# Quark Masses in Supersymmetric $SU(3)_{color} \otimes SU(3)_W \otimes U(1)_X$ Model with Discrete $T^{/}$ flavor symmetry

Sutapa Sen

*Department of Physics, Christ Church P.G. College, Kanpur 208001, India*

## Abstract

We propose a supersymmetric model of quark flavor symmetry based on $SU(3)_{color} \otimes SU(3)_W \otimes U(1)_X \otimes T^{/}$ where $T^{/}$ ( binary tetrahedral group) is a discrete flavor symmetry group. The third (heavy) generation of quarks are assigned to singlet representation of $T^{/}$ while light quark generations are treated differently and assigned to doublet representation of $T^{/}$ . The model generates masses for heavy quarks as in 3-3-1 case. The quark mass terms for the second generation are obtained when $T^{/}$ symmetry is broken at leading order and for first generation with small higher order corrections .The flavor-changing neutral current (FCNC) effects are suppressed at leading order for $3 \rightarrow 2$ quark transitions due to $T^{/}$ breaking which is an attractive feature of the 3-3-1 $\otimes T^{/}$ model.

## 1. Introduction

Recently, experimental data on neutrino mixing matrix [1] has been interpreted successfully by introducing a discrete flavor symmetry group [2] to the structure of the flavor gauge group. The models based on discrete $A_4$ symmetry group [3-5] offer predictions for tri-bimaximal neutrino mass matrix [6] in agreement with present data for the lepton sector. The supersymmetric extension of Standard Model gauge group (MSSM) with $A_4$ discrete flavor symmetry has been recently considered [4] for both lepton and quark sectors. This has limited success for the quark sector[5].A different approach is to consider three generations of lepton and quarks with $T^{/}$ or $SL_2$ ($F_3$), the binary tetrahedral symmetry group [8] which is the double covering of $A_4$ .This has been considered by several authors in framework of the Standard model (SM) [7,8], in SU(5) framework [9] and in SUSY MSSM model [10] for tri-bimaximal neutrino mixing and realistic patterns for quark masses. In addition to the masses and mixing in quark and lepton sectors, the discrete flavor symmetry can also throw light on protection from flavor-changing neutral currents (FCNC) which are generally enhanced in multi-Higgs supersymmetric (SUSY) models [11] thus ruling out such models.

The $SU(3)_c\otimes SU(3)_L\otimes U(1)_X$ (3-3-1) gauge group [12] with SUSY extension[13,14] offers an interesting framework for addition of discrete flavor $T^{/}$ symmetry group. for investigation of quark masses and FCNC and is the basic motivation of the present work. The neutrino mass mixing matrix with heavy lepton has been recently considered in the 3-3-1 model with $A_4$ symmetry [15] .

In this work, we consider the M331SM model [16] with addition of discrete $T^{/}$ flavor group .This provides a natural framework for superpotential with third generation treated differently from the two light generations of quarks. In the lepton sector, the $T^{/}$ group maintains the results for $A_4$ group .We do not consider this sector in the present work.

The main results include quark masses and suppression of FCNC in the SUSY 3-3-1 model with additional discrete $T^{/}$ symmetry. The quark mass terms in the third generation are $T^{/}$ invariant and allow heavy top, bottom and exotic quark masses due to 3-3-1 Yukawa mass terms with additional Higgs scalars of SUSY 3-3-1 [12-14,16] .These are distinct from the 3-3-1 scalars which generate masses for the first and second generations. The light fermions are assigned to ($2_3$ ) doublet representations of $T^{/}$ which is broken below cutoff scale $\Lambda$ by 3-3-1 singlet scalars (flavon).The light quarks (u,d) do not attain mass at leading order. This is corrected by small effects coming from higher order corrections [10].

The corrected Yukawa Lagrangian allows FCNC effects at tree level which are suppressed by vacuum expectation values (VEVs) gained by 3-3-1 singlet flavon due to $T^{/}$ symmetry breaking. In Section 2, we present an outline of the SUSY 3-3-1 model.. Section 3 deals with discrete flavor symmetry $T^{/}$, the quark mass matrices at leading order and with corrections. Section 4 deals with suppression of FCNC effects and constraints on Higgs decays to fermions . Section 5 deals with results and conclusion

## 2. Outline of SUSY3-3-1 model (M331SM) with $T^{/}$ symmetry

The non-SUSY $SU(3)_C \otimes SU(3)_L \otimes U(1)_X$ model [Pleitez-Tonasse version] in ref.[12] is considered as embedded in $SU(4)_{PS} \otimes SU(4)_W$ gauge group [17] with symmetry-breaking pattern as follows:

$$SU(4)_{PS}\otimes SU(4)_W \xrightarrow{M_{GUT}} SU(3)_C\otimes SU(3)_L\otimes U(1)_{B-L}\otimes U(1)_\psi$$

$$\xrightarrow{M_R} SU(3)_C\otimes SU(3)_L\otimes U(!)_X$$

$$\xrightarrow{\langle\chi\rangle} SU(3)_C\otimes SU(2)_L\otimes U(!)_{X'}\otimes U(1)_X$$

$$\xrightarrow{\langle\chi\rangle} SU(3)_C\otimes SU(2)_L\otimes U(!)_Y \qquad (1)$$

The $SU(4)_{PS}$ generator $T^{15}{}_{PS} = \sqrt{\frac{3}{2}}\,(B\text{-}L)/2 = \sqrt{\frac{3}{2}}\,\mathrm{diag}\,(1/6,1/6,1/6,-1/2)$;

$SU(4)_W$ generator $T^{15}{}_W = \frac{1}{\sqrt{6}} Y_\psi = \frac{1}{\sqrt{6}}\,\mathrm{diag}\,(1/2,1/2,1/2,-3/2)$ .

- Symmetry-breaking at first stage is at GUT scale with SU(4)→ SU(3) ⊗U(!) [17] for both $SU(4)_{PS}$ and $SU(4)_W$ by considering Higgs scalars $H(15_{PS},1_W)$ and $H(1_{PS},15_W)$.
- At second stage, $3_C$-$3_L$-$1_\psi$ -$1_{B-L}$ → $3_C$-$3_L$-$1_X$ at $M_R$ scale .The neutral singlet bilepton $\phi_0$ ($1_C$,$1_L$, -1,1) is a scalar field first proposed in [17] ($Y_\psi$ = -1,B-L =2) which gains vacuum expectation value (VEV) , $<\phi_0> = w$. The $U(1)_X$ charge $X = Y_\psi + (B\text{-}L)/2 = \sqrt{6}\, T^{15}{}_W + I_4\,(B\text{-}L)/2$ (2)
- Further breaking $3_C$-$3_L$-$1_X$ → $3_C$-$2_L$-$1_{X'}$-$1_X$ is obtained by introducing 3-3-1 scalar $\chi(1_C,3_L,-1)$, which gains VEV $<\chi^0> = V$ at TeV scale, with charge $X' = \sqrt{3}\, T^8{}_W$ where the SU(4) generator $T^8{}_W = \frac{1}{\sqrt{3}}\,\mathrm{diag}\,(1/2, 1/2, -1, 0)$ .

The hypercharge $Y = 2(X-X')$ so that electric charge operator

$$Q/e = T^3{}_L + Y/2 = T^3{}_L + X - X' = T^3{}_L - \sqrt{3}\, T^8{}_W + \sqrt{6}\, T^{15}{}_W + I_4\,(B\text{-}L)/2 \qquad (3)$$

The linear combination of diagonal generators $X' = -T^3{}_R + Y_\psi$ is preserved by $\phi_0$ scalar along with hypercharge Y. Additional scalars $\eta(1_C,3_L,0)$ , $\rho(1_C,3_L,1)$ break

$SU(2)_L \otimes U(!)_Y \rightarrow U(1)em$ and gain VEV's at electroweak scale

$$<\eta^0> = v_u, \quad < \rho^0> = v_d. \tag{4}$$

## 2.1 Scalar sector of SUSY 3-3-1 model.

We consider a supersymmetric version of the 3-3-1 model as follows.

**Higgs triplets:** The Higgs scalars include weak isospin doublet and a singlet of $SU(2)_L$

$$\eta = \begin{pmatrix} \eta^0 \\ \eta_1^- \\ \eta_2^+ \end{pmatrix} \sim (1,3,0); \rho = \begin{pmatrix} \rho^+ \\ \rho^0 \\ \rho^{++} \end{pmatrix} \sim (1,3,1); \chi = \begin{pmatrix} \chi^- \\ \chi^{--} \\ \chi 0 \end{pmatrix} \sim (1,3,-1) \tag{5}$$

The vacuum expectation values which are nonzero include

$$< \eta^0> = v_u, \ <\rho^0> = v_d, \ < \chi^0>= V \tag{6}$$

We introduce chiral superfields $\hat{\phi}$ by extending the particle content to include squarks, sleptons and higgsinos .The superpartner for a given particle $f$ is $\widetilde{f}$ .

For SUSY 3-3-1 model, the cancellation of chiral anomalies generated by superpartners of scalar fields requires three extra scalars and higgsinos [13].

$$\eta' = \begin{pmatrix} \eta'^0 \\ \eta_1'^+ \\ \eta_2'^- \end{pmatrix} \sim (1,3^*,0); \rho' = \begin{pmatrix} \rho'^- \\ \rho'^0 \\ \rho'^{--} \end{pmatrix} \sim (1,3^*,-1); \chi' = \begin{pmatrix} \chi'^+ \\ \chi'^{++} \\ \chi'^0 \end{pmatrix} \sim (1,3^*,+1) \tag{7}$$

The vacuum expectation values which are nonzero include

$$<\eta'^0> = v_u', \ <\rho'^0> =v_d', \ <\chi'^0> = V' \tag{8}$$

The extended electroweak sector can be compared with minimal supersymmetric standard model (MSSM) and $ME_6SSM$ [19,20] .The main difference is that M331SM contains two sets of MSSM-like Higgs doublet scalars with hypercharge $Y= 2( X – X')$

*(i)* $(\eta^0, \eta_1^-), (\rho^+, \rho^0)$ *; Y = -1,+1;*

*(ii)* $(\rho'^-, \rho'^0), (\eta^{0'}, \eta_1^{+'})$*; Y = -1,+1*

Set (i) scalars couple only to generations 1,2 of light quarks and correspond to a second set of MSSM doublets ($h_1$,$h_2$). Set (ii) scalars couple only to the third generation quarks which are singlets of flavor group $T'$.

Since $X' = -T^3_R + Y_\psi$ , the scalar fields of $3_C$-$2_L$-$1_{X'}$-$1_X$ can be considered as representations of gauge group $SU(3)_C \otimes G$ where $G = SU(2)_L \otimes U(1)_{T^3_R} \otimes U(1)_\psi$.

The MSSM doublet $h_1$ {as in $ME_6SSM$ [19,20] Higgs doublet $h_1 \sim (2_L,-1/2, -1)$} is embedded as $(\rho^{-'}, \rho^{0'})$ in 3-3-1 case. The Higgs doublet $h_2$ in 3-3-1 transforms under G as $(\eta_1^{+'}, \eta^{0'}) \sim (2_L, 1/2 ,0)$ which is different from $ME_6SSM$ doublet $h_2 \sim (2_L,1/2,-1)$.

The exotic scalars in 3-3-1 include $SU(2)_L$ singlet components $\eta_2^+$ $(\eta_2'^-) \sim (1_L, \pm 1,0)$, $T_R^3 = \pm 1$, while $\rho^{++}(\rho'^{--}) \sim (1_L, \pm 2, \pm 1)$ are $T_R^3 = \pm 2$ scalars. The scalar $\chi'$ $(\chi)$ is assigned to $10_S$ plet representation of $SU(4)_W$. The neutral component $\chi'^0(\chi^0) \sim (1_L,0,\pm 1)$.

## 2.2 Physical Higgs fields for M331SM

To obtain physical scalar Higgses, we introduce the expansion of neutral scalar fields,

$X_i^0 = \frac{1}{\sqrt{2}}(v_{X_i} + \xi_{X_i} + i\zeta_{X_i})$ where the vacuum expectation values include

$<\eta^0> = v_u$, $<\rho^0> = v_d$. $<\chi^0> = V$; $<\eta'^0> = v'_u$, $<\rho'^0> = v'_d$, $<\chi'^0> = V'$

$$tan\,\beta = \frac{v_d}{v_u},\; tan\,\beta' = \frac{v'_d}{v'_u} \qquad (9)$$

The mass spectrum for physical Higgs fields have been considered in ref.{14-16]. The main results are

- The neutral scalar sector includes three pairs of massive physical fields. In the basis of $(\xi_\eta, \xi_\rho)$, $(\xi_\eta^/, \xi_\rho^/)$, $(\xi_\chi, \xi_\chi^/)$ these are given by $(H_1^0, h_1^0)$, $(H_2^0, h_2^0)$, $(H_3^0, h_3^0)$

$$H_1^0 = \cos\alpha_1 \xi_\eta + \sin\alpha_1\, \xi_\rho \ ; \quad h_1^0 = -\sin\alpha_1 \xi_\eta + \cos\alpha_1\, \xi_\rho$$

$$H_2^0 = \cos\alpha_2 \xi_\eta^/ + \sin\alpha_2\, \xi_\rho^/ \ ; \quad h_2^0 = -\sin\alpha_2\, \xi_\eta^/ + \cos\alpha_2\, \xi_\rho^/$$

$$H_3^0 = \cos\alpha_3 \xi_\chi + \sin\alpha_3\, \xi_\chi^/ \ ; \quad h_3^0 = -\sin\alpha_1 \xi_\chi + \cos\alpha_3\, \xi_\chi^/$$

- The physical neutral pseudoscalar fields include two massless Goldstone bosons $G_1^0, G_2^0$ and four massive pseudoscalars $A_1, A_{2,} A_3, A_4$.

- In the singly charged sector, two charged Goldstone bosons $G_1^\pm$,$G_3^\pm$ are obtained with six massive scalars $H_1^\pm$, $H_2^\pm$, $h_2^\pm$, $H_3^\pm$, $H_4^\pm$ and $h_4^\pm$

- The spectrum of doubly charged scalars includes one doubly charged Goldstone boson $G^{\pm\pm}$ and three massive physical fields .$H^{\pm\pm}$, $H^{/\pm\pm}$ and $h^{\pm\pm}$

## 2.3 Matter content of SUSY 3-3-1 model

The matter content of M331SM includes anomaly-free three generations of quarks and leptons along with their superpartners.

**Matter multiplets:**

Lepton $\psi_{\alpha L} = \begin{pmatrix} \nu_{l\alpha} \\ l_\alpha \\ P_\alpha^+ \end{pmatrix} \sim (1_C, 3_L, 0)$ , $l_\alpha = e, \mu, \tau; P_\alpha = P_e, P_\mu, P_\tau$

Quark $Q_i = \begin{pmatrix} d_i \\ u_i \\ D_i \end{pmatrix} \sim \left(3_C, 3^*_L, -\frac{1}{3}\right), i = 1,2$ ; $Q_3 = \begin{pmatrix} t \\ b \\ T \end{pmatrix} \sim \left(3_C, 3_L, \frac{2}{3}\right)$

The singlet leptons are

$l_R^C \sim (1_C, 1_L, -1); P_R^C \sim (1_C, 1_L, 1)); l = e, \mu, \tau; P = P_e, P_\mu, P_\tau.$

The $SU(3)_L$ singlet quarks include

$$u_{Ri}^C \sim \left(3_C^*, 1_L, -\frac{2}{3}\right); d_{Ri}^C \sim \left(3_C^*, 1_L, \frac{1}{3}\right); D_{Ri}^C \sim \left(3_C^*, 1_L, \frac{4}{3}\right);$$

$$t_R^C \sim \left(3_C^*, 1_L, -\frac{2}{3}\right); b_R^C \sim \left(3_C^*, 1_L, \frac{1}{3}\right); T_R^C \sim \left(3_C^*, 1_L, -\frac{5}{3}\right). \quad (10)$$

The discrete flavor symmetry group $T^/$ is added to the gauge group such that the third generation of heavy quarks transforms trivially under this group..

## 3. Discrete flavor symmetry $T^/$ and flavons

There are 24 elements in representations of $T^/$ group which include two types , that of $A_4$ ( $1_1$ , $1_2$ , $1_3$ and 3) and three additional doublets,( $2_1$,$2_2$ and $2_3$) which cannot be decomposed into representations of $A_4$ (not a subgroup of $T^/$) .We follow ref.[9,10] for the pattern of symmetry-breaking of flavor group $T^/$. Above the cutoff scale $\Lambda$, $T^/$ symmetry is exact. At leading order, $T^/$ breaks down to subgroup $G_T$ generated by T, in charged fermion sector. Various 3-3-1 singlet flavons develop vacuum expectation values (VEVs)

$$T^/ \rightarrow G_T : <\phi_T(3)> = (v_T, 0,0)\ ,\ <\xi^{//}(1_3)> = 0\ ,\ <\psi(2_2)> = (v_1,0)\ ,$$

$$T^/ \rightarrow \text{nothing:} \qquad <\psi^/(2_1)> = (v_1^/, v_1^/) \quad (11)$$

The Kronecker products for the irreducible representations of $T^/$ listed in ref. [8-10] have been used to obtain the Yukawa mass terms for M331SM . At higher order, operators which depend on the fields $\phi_T$, $\xi^{//}$, $\psi$ and $\psi^/$ only and leave $T^/$ invariant after spontaneous symmetry-breaking[10] predict same textures for up and down mass matrices. A small correction is thus introduced to VEVs of these fields which are shifted as [10]

$$<\Phi_T> = ( v_T + \delta v_{T1}, \delta v_{T2}, \delta v_{T3}) , \; < \psi> = (v_1+ \delta v_1, \delta v_2), <\xi^{//}> = \delta u^{//}. \quad (12)$$

We consider $\delta v_T = \delta v_{T1} = \delta v_{T2} = \delta v_{T3}$ , $\delta u^{//} = 0$.The consequences of higher order corrections are given in eqn.(15-16).The representations of $T^{/}$ and assignment of scalar fields and flavons to these are given in Table 1.

The addition of the discrete symmetry group $T^{/}$ to M331SM modifies the superpotential

$$W = W_Q + W_L + W_T^{/} \quad (13)$$

where $W_Q$ , $W_L$ denote the mass terms for quark and lepton while $W_T^{/}$ contains terms for vacuum alignment. The cut-off scale above which $T^{/}$ is exact is given by Λ. The superpotential terms include new supermultiplet (flavon) or driving fields which are assigned to $T^{/}$ representations as $\Phi_T(3)$, $\xi^{//}(1_3)$, $\psi(2_2)$ and $\psi^{/}(2_1)$. At leading order $\Phi_T(3)$, $\xi^{//}(1_3)$, $\psi(2_2)$ and $\psi^{/}(2_1)$ gain VEV's by symmetry-breaking [10]

$$<\phi_T(3)> = (v_T, 0,0) , \; <\xi^{//} (1_3)> = 0 , <\psi(2_2)> = ( v_1,0) , \; <\psi^{/} (2_1)> = ( v_1^{/},v_1^{/}) \quad (14)$$

### 3.1 Corrections to quark masses and mixing angles

The trilinear mass terms in superpotential $W_Q$ for quark sector include (i,j = 1,2)

$$\begin{aligned} W_3 = \; & k^b ( b^c Q_{3L}) \rho^{/} + k^t ( t^c Q_{3L}) \eta^{/} + k^T ( T^c Q_{3L}) \chi^{/} + f_1 \rho\eta\chi + f_1^{/}\rho'\eta'\chi' + \\ & \Sigma_{i j} (\Phi_T/\Lambda)\{ k^u{}_{i j} (u^c{}_{j R} Q_{iL}) \rho + k^d{}_{i j} (d^c{}_{j R} Q_{iL}) \eta\} + \Sigma_{i ,j} (\psi^{/} /\Lambda) k^D{}_{i j} (D^c{}_{j R} Q_{iL}) \chi + \\ & \Sigma_{i j} ( \xi^{//} /\Lambda)\{ k^u{}_{i j} (u^c{}_{j R} Q_{iL}) \rho + k^d{}_{i j} (d^c{}_{j R} Q_{iL}) \eta\}+ \\ & \Sigma_i (\psi /\Lambda)\{k^t{}_i(t^c Q_{iL})\rho+ k^b{}_i (b^c Q_{iL})\eta\}+ \Sigma_j (\psi/\Lambda)\{k^u{}_j (u^c{}_{jR}Q_{3L})\eta^{/}+ k^d{}_i (d^c{}_{jR} Q_{3L})\rho^{/}\} \end{aligned} \quad (15)$$

We note that two sets of Higgs doublets can couple differently to heavy and light quarks which is different from the MSSM model ref [10] with one set of ( $h_{u,}$ ,$h_d$) Higgs scalars. At leading order, the mass matrices for up and down quarks

$$M_u = \begin{pmatrix} 0 & 0 & 0 \\ 0 & k_{22}^{u}\frac{v_T}{\Lambda}v_d & k_2{}^{u}\frac{v_1}{\Lambda}v_u' \\ 0 & k_2{}^{t}\frac{v_1}{\Lambda}v_d & k^t v_u' \end{pmatrix} \qquad M_d = \begin{pmatrix} 0 & 0 & 0 \\ 0 & k_{22}^{d}\frac{v_T}{\Lambda}v_u & k_2{}^{d}\frac{v_1}{\Lambda}v_d' \\ 0 & k_2{}^{b}\frac{v_1}{\Lambda}v_u & k^b v_d' \end{pmatrix} \qquad (16)$$

The quark masses at leading order are

$$m_u = 0,\ m_d = 0\ ;\ m_c = k_{22}^{u}\frac{v_T}{\Lambda}v_d\ ;\ \ m_s = k_{22}^{d}\frac{v_T}{\Lambda}v_u\ ;\ m_t = k^t v_u{}'\ ;\ m_b = k^b v_d{}' \qquad (17)$$

The exotic quarks gain masses $m_T = k^T V'$ ; $m_D = \begin{pmatrix} k_{11}^D & k_{12}^D \\ k_{21}^D & k_{22}^D \end{pmatrix} V v_1{}'$ (18)

From eqn .(16), $V_{us} = 0$, $V_{ub} = 0$ so that the model requires additional mass corrections. The advantage of additional suppression factors due to VEV's of flavon $(\Phi_T/\Lambda) \sim \lambda^2$, $(\psi/\Lambda) \sim \lambda^2$ in the superpotential $W_Q$ is that FCNC at tree level for Higgs couplings to fermions are suppressed by a factor $\lambda^2$ for 3→2 but is zero for 3→1 transition of quarks.

As discussed in ref [10], the main corrections to quark masses are considered by modifying the VEV's of the flavons .The terms in the superpotential responsible for $T'$ symmetry-breaking include supermultiplets describing flavon fields $\Phi_T$, $\xi''$, $\psi$ and $\psi'$ which develop VEV's. We follow the method outlined in ref [10] and modify the VEV's which are shifted to

$$\langle\Phi_T\rangle = (v_T + \delta v_{T1},\ \delta v_{T2},\ \delta v_{T3})\ ,\ \langle\psi\rangle = (v_1 + \delta v_1,\ \delta v_2)\ ;\ \langle\xi''\rangle = 0 \qquad (19)$$

The corrected mass matrices are similar to that in ref[10] but now in a three-generation SUSY 3-3-1 model

$$M_u = \begin{pmatrix} ik_{11}^u \frac{\delta v_T}{\Lambda} v_d & \frac{(1-i)}{2} k_{12}^u \frac{\delta v_T}{\Lambda} v_d & k_1^t \frac{\delta v_2}{\Lambda} v_d \\ \frac{(1-i)}{2} k_{21}^u \frac{\delta v_T}{\Lambda} v_d & k_{22}^u \frac{v_T}{\Lambda} v_d & k_2^u \frac{v_1}{\Lambda} v_u' \\ k_1^u \frac{\delta v_2}{\Lambda} v_u' & k_2^t \frac{v_1}{\Lambda} v_d & k^t v_u' \end{pmatrix} \quad (20)$$

$$M_d = \begin{pmatrix} ik_{11}^d \frac{\delta v_T}{\Lambda} v_u & \frac{(1-i)}{2} k_{12}^d \frac{\delta v_T}{\Lambda} v_u & k_1^b \frac{\delta v_2}{\Lambda} v_u \\ \frac{(1-i)}{2} k_{21}^d \frac{\delta v_T}{\Lambda} v_u & k_{22}^d \frac{v_T}{\Lambda} v_u & k_2^d \frac{v_1}{\Lambda} v_d' \\ k_1^d \frac{\delta v_2}{\Lambda} v_d' & k_2^b \frac{v_1}{\Lambda} v_u & k^b v_d' \end{pmatrix} \quad (21)$$

By diagonalizing the above matrices,

$$m_u : m_c : m_t = k_{11}^u \delta v_T v_d / \Lambda : k_{22}^u v_T v_d / \Lambda : k^t v_u'$$

$$m_d : m_s : m_b = k_{11}^d \delta v_T v_u / \Lambda : k_{22}^d v_T v_u / \Lambda : k^b v_d' \quad (22)$$

The VEVs $v_d^{/}$, $v_u^{/}$ give heavy masses to $m_b$, $m_t$ at electroweak scale.

$$m_t = k^t v_u^{/} ; m_b = k^b v_d^{/} \quad (23)$$

The observed quark masses obey the relations [9]

$$m_u : m_c : m_t = \varepsilon_u^2 : \varepsilon_u : 1, \qquad m_d : m_s : m_b = \varepsilon_d^2 : \varepsilon_d : 1$$

where $\varepsilon_u = 0.005$, $\varepsilon_d = 0.05$. (24)

From eqn.(22-24), we obtain the relations for Yukawa parameters

$$m_t / m_b = 100 \tan \beta k^u{}_{11} / k^d{}_{11} = 10 \tan \beta \; k^u{}_{22} / k^d{}_{22} \quad (25)$$

$$m_u / m_c = k_{11}^u / k_{22}^u (\delta v_T / v_T); m_d / m_s = k_{11}^d / k_{22}^d (\delta v_T / v_T) \quad (26)$$

## 4. Suppression of FCNC and constraints on Higgs decays

Additional suppression factor for FCNC transitions due to enhanced Higgs couplings to fermions at tree level ( for multi-Higgs models) is possible by choosing [10]

$VEV/\Lambda = \lambda^2$ .Here VEV is the non-vanishing VEV for flavons ($\Phi_T$, $\xi^{//}$, $\psi$ and $\psi^{/}$) while $\Lambda$ is the cut-off scale for $T^{/}$ group. This suppresses the predictions of 3-3-1 model for $t \rightarrow c\, H$ and $b \rightarrow s\, H$ decay rates at tree level by a factor $v_1/\Lambda \sim \lambda^2$ .

$$g(tcH_2^0) = k^u{}_2\, v_u^{/}\ (v_1/\Lambda) \qquad (27)$$

From correction factors due to $\delta v_T$, $\delta v_2$ in eqn.(15) additional suppression of FCNC transitions $t \rightarrow u\, H$ , $b \rightarrow d\, H$ is obtained at tree level

$$g(tuH_2^0) = k^u{}_1\, v_u^{/}\, (\delta v_2/\Lambda) \qquad (28)$$

The mass terms in the superpotential $W_Q$ in eqn.(15) are significantly different from MSSM since the heavy scalar triplets [($\rho^{0\,/}$,$\rho^{-\,/}$ , $\rho^{/\,--}$), ($\eta_1^{+\,/}$, $\eta^{0\,/}$, $\eta_2^{/\,-}$)] can couple only to the third generation quarks. The neutral components $\rho^{0/}$, $\eta^{0/}$ are mixed to obtain physical Higgses $h_2^0$, $H_2^0$. These scalars contribute to the decay $H^0 \rightarrow b\ \bar{b}$ at tree level. The light quarks (u,d) ,(c,s) couple only to the scalars ($\eta^0$, $\eta_1^-$), ($\rho^+$,$\rho^0$) . These scalars are mixed to obtain physical states $h_1$,$H_1$ which do not have tree level $b\ \bar{b}$ couplings.. Since both types of scalars can be produced in $e^+ e^- \rightarrow Z\, H$, this will constrain the $H \rightarrow b\ \bar{b}$ decays for $h_1$, $H_1$ scalars. This feature has important phenomenological significance for Higgs searches at LHC as has been pointed out recently [20]

## 5. Conclusions

The addition of discrete flavor symmetry group $T^{/}$ (double tetrahedral group) to M331SM model is shown to improve considerably the quark masses and Yukawa parameters .This work is distinct from the SUSY MSSM model with $T^{/}$ symmetry in [10] which has been considered for the vacuum alignment problem, both at leading order and with small higher order corrections. The main differences include

(i) two sets of Higgs doublets which couple differently to the third and first two generations of quarks. The couplings of ( $h_2$ , $H_2$) to third generation (t, b, T) quarks are $T^{/}$ invariant while those for ($h_1$,$H_1$) to light quarks are constrained by $T^{/}$ symmetry.

(ii) The Yukawa parameters are already separated for the third and first two generations. which transform differently under 3-3-1 symmetry.

(iii) New set of relations are obtained from eqn.(25-26) for Yukawa parameters due to $T^{/}$ symmetry.

(iv) The $U(1)_{FN}$ Froggatt and Nielson flavor symmetry [10] is not required in this model.

The M331SM model with discrete $T^{/}$ symmetry group provides a dynamical explanation to the problem of mass hierarchy in the quark sector .A second attractive consequence is the suppression of FCNC in this multi-Higgs SUSY model by factors $v_1/\Lambda$, $\delta v_2/\Lambda$ due to spontaneous breaking of $T^{/}$ symmetry. A detailed numerical analysis of Higgs masses and decays in M331SM would throw interesting light on the Higgs phenomenology at LHC and future colliders.

## References

[1] For a recent review, see M. C.Gonzalez-Garcia and M. Maltoni, arXiv:0704.1800[hep-ph]

[2] E. Ma, arXiv: hep-ph/0409075; E. Ma, arXiv: 0705.0327[hep-ph]

[3] E.Ma and G.Rajasekaran, Phys.Rev.**D64**,113012 (2001);K .S. Babu, E .Ma and J.W.F.Valle, Phys.Rev.Lett.**B552**, 207(2003)

[4] F.Bazzocchi, S.Kaneko and S.Morisi,arXiv:07073032 v1[hep-ph];E.Ma, Phys.Rev.**D 73**,057304 (2006).

[5] G.Altarelli and F.Feruglio, Nucl.Phys.**B720** ,64(2005), hep-ph/0504165. G.Altarelli and F.Feruglio, Nucl.Phys.**B741** ,215(2006), hep-ph/0512103. G.Altarelli and F.Feruglio, arXiv:hep-ph/0610165

[6] P .F. Harrison, D.H. Perkins and W. G. Scott, Phys. Lett. **B 530**, 167(2002)

[7] P.H.Frampton and T.W.Klephart, Int.J.Mod.Phys.**A10**,4689 (1995); P.H.Frampton and T.W.Klephart ,Phys.Rev.**D64**, 086007(2001);P.D.Carr and P.H.Frampton, arXiv: hep-ph/0701034; P.H.Frampton and T.W.Klephart , arXiv:0706.1186v2 [hep-ph]

[8] A .Aranda, C.D.Carone and R .F. Lebed, Phys.Lett.**B 474**, 170(2004); A.Aranda, C.D.Carone and R.F.Lebed, Phys.Rev.**D62**, 016009(2000).

[9] M .C .Chen and K .T .Mahanthappa ,arXiv:0705.071v3[hep-ph]

[10] F.Feruglio , C.Hagedorn ,Y. Lin and L.Merlo, Nucl. Phys. **B 775**,120(2007) **.** arXiv **:** 0702194v1[hep-ph]

[11] N. Escudero, C. Munoz and A.M. Teixeira, Phys.Rev.**D73**, 055015(2006)

[12] F.Pisano and V. Pleitez, Phys.Rev.**D46**, 410(1992); R. Foot, O. F. Hernandez,

F.Pisano and V.Pleitez, Phys.Rev.**D47**,4158 (1993); N.T.Anh, N. Anh Ky and H.N.Long, Int.J.Mod.Phys.**A16**,541(2001);P.Frampton,Phys.Rev.Lett.**69**,2889(1992);

. V.Pleitez and M.D.Tonasse, Phys. Rev.**D48**,(1993) 2353.

M. D. Tonasse, Phys.Lett.**B381**, (1996)191.J.E.Cieza Montalvo and M. D. Tonasse , Nucl.Phys.**B623**,(2002)325.M.Capdequi-Peyranere and M. C. Rodriguez, [arXiv: hep-ph/0103013].Phys.Rev.**D65**,(2002)035001

[13] T. V .Duong and E. Ma, Phys. Lett. **B 316**, 307 (1993); H. N. Long and P .B .Pal, Mod .Phys .Lett .**A 13**, 2355 (1998)

[14] M. Capdequi- Peyranere and M. C. Rodriguez ,[arXiv: hep-ph/0103013] .Phys.Rev.**D65**,(2002)035001. J. C. Montero, V. Pleitez and M .C. Rodriguez, Phys .Rev .**D65**, 035006 (2002);

J.C.Montero, V. Pleitez , M. C .Rodriguez, Phys .Rev .**D65,** 095008(2002)

J.C.Montero, V. Pleitez and M.C .Rodriguez, Phys.Rev.**D70** ,075004.(2004)

Sutapa Sen and A Dixit [ arXiv: hep-ph /(0503078)]

[15] F. Yin , arXiv: 07043827v1[hep-ph]

[16] M. C. Rodriguez, [arXiv:hep-ph/0510333]

Sutapa. Sen and A. Dixit,[arXiv:hep-ph/0510393]

[17] Sutapa Sen and A. Dixit, Phys.Rev.**D71**,035009 (2005)

[18] R.How and S.F.King,arXiv.org: 0708.1451[hep-ph]

[19] R.How and S.F.King,arXiv.org: 0705.0301[hep-ph]

[20] R.Dermisek and J.F.Gunion, arXiv.org :0709.2269v1[hep-ph]

TABLE 1: The transformation rules of the fields under 3-3-1 and $T^{/}$ symmetries .The fermions include $D_q(Q_1,Q_2)_L$ ,$D_u^c$ $(u^c,c^c)$, $D_d^c$( $d^c,s^c$), $D_D^c(D_1^c,D_2^c)$ as doublets of $T^{/}$ symmetry The $L_\alpha$ include three generations of left-handed leptons in eqn.(10) as triplet representation of $T^{/}$ symmetry.

The 3-3-1 scalars are singlets of $T^{/}$ while the driving fields of $T^{/}$ are singlets of 3-3-1 symmetry.

| Field | $3_C$-$3_L$-$1_X$ | $T^{/}$ | Field | $3_C$-$3_L$-$1_X$ | $T^{/}$ | Field | $3_C$-$3_L$-$1_X$ | Field | $T^{/}$ |
|---|---|---|---|---|---|---|---|---|---|
| $D_q$ | (3,3*,-1/3) | $2_3$ | $L_\alpha$ | (1,3,0) | 3 | $\eta$ | (1,3,0) | $\varphi_T$ | 3 |
| $Q_3$ | (3,3,2/3) | $1_1$ | $\tau^c$ | (1,1,1) | $1_1$ | $\rho$ | (1,3,1) | $\xi^{//}$ | $1_3$ |
| $D_u^c$ | (3*,1,-2/3) | $2_3$ | $\mu^c$ | (1,1,1) | $1_2$ | $\chi$ | (1,3,-1) | $\psi$ | $2_1$ |
| $D_d^c$ | (3*,1,1/3) | $2_3$ | $e^c$ | (1,1,1) | $1_3$ | $\eta^{/}$ | (1,3*,0) | $\psi^{/}$ | $2_2$ |
| $D_D^c$ | (3*,1,4/3) | $2_2$ | $P_3^c$ | (1,1,-1) | $1_1$ | $\rho^{/}$ | (1,3*,-1) | | |
| $t^c$ | (3*,1,-2/3) | $1_1$ | $P_2^c$ | (1,1,-1) | $1_2$ | $\chi^{/}$ | (1,3*,1) | | |
| $b^c$ | (3*,1,1/3) | $1_1$ | $P_1^c$ | (1,1,-1) | $1_3$ | $\varphi_0$ | (1,1,0) | | |
| $T^c$ | (3*,1,-5/3) | $1_1$ | $\nu_{R\alpha}^c$ | (1,1,0) | $1_1$ | | | | |